\documentclass[twocolumn,twoside,slac]{revtex4}
\usepackage{graphicx}
\usepackage{fancyhdr}
\begin{document}

\title{Improving the Security and Performance of the BaBar Detector Controls System}

%

\author{K. D. Kotturi}
\affiliation{For the BaBar Computing Group, SLAC, Stanford, CA 94025, USA}

\begin{abstract}
It starts out innocently enough - users want to monitor Online data and so run
their own copies of the detector control GUIs in their offices and at home. But
over time, the number of processes making requests for values to display on
GUIs, webpages and stripcharts can grow, and affect the performance of an
Input/Output Controller (IOC) such that it is unable to respond to requests
from requests critical to data-taking. At worst, an IOC can hang, its CPU
having been allocated 100\% to responding to network requests.

For the BaBar Online Detector Control System, we were able to eliminate this
problem and make great gains in security by moving all of the IOCs to a
non-routed, virtual LAN and by enlisting a workstation with two network
interface cards to act as the interface between the virtual LAN and the public
BaBar network. On the interface machine, we run the Experimental Physics
Industrial Control System (EPICS) Channel Access (CA) gateway software
(originating from Advanced Photon Source). This software accepts as inputs, all the channels which
are loaded into the EPICS databases on all the IOCs. It polls them to update
its copy of the values. It answers requests from applications by sending them
the currently cached value.

We adopted the requirement that data-taking would be independent of the
gateway, so that, in the event of a gateway failure, data-taking would be
uninterrupted. In this way, we avoided introducing any new risk elements to
data-taking.  Security rules already in use by the IOC were propagated to the
gateway's own security rules and the security of the IOCs themselves was
improved by removing them from the public BaBar network.

\end{abstract}

\maketitle

\thispagestyle{fancy}


\section{Introduction}

This paper describes the motivation behind and implementation of 
a CA Gateway \cite{gw-ref} in the BaBar Online Detector Controls System. 

\section{Motivation}

With the IOCs on the BaBar public network, there were no limits on the
number of clients (aside from the limit of 384 ssh sessions) that could connect and request values. 
These clients could run on any of the 63 workstations on the BaBar
public network. Remote access to the BaBar public network was possible
via ssh.

As a result, the number of
open file descriptors on an IOC could exceed the 150 limit (already increased
significantly from the default 50) and the CPU usage of an IOC could go 
as high as 100

\section{Implementation}

\begin{figure}
\includegraphics[width=65mm]{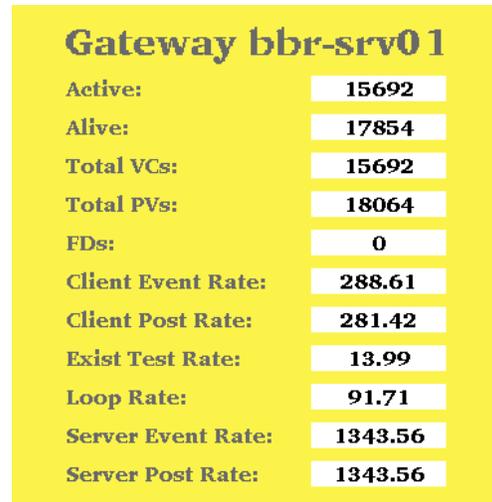}
\caption{CA gateway statistics.}
\label{stats}
\end{figure}

\begin{figure}
\includegraphics[width=65mm]{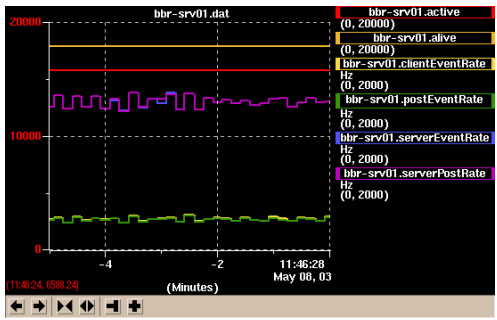}
\caption{CA gateway statistics as a function of time.}
\label{graph}
\end{figure}

The IOCs were moved to a private network and the four servers running EPICS
client software that were essential for data-taking were given a 
secondary network interface so that they could access the private network. 
CA gateway software was installed on one of these servers, "bbr-srv01". 
Statistics of the gateway are displayed instantaneously in Figure~\ref{stats} and over
time in Figure~\ref{graph}. 

The gateway acts as a client to the IOCs, requesting channels and receiving replies at the rate of approximately 285 channels per sec (Figure~\ref{stats}). 

The gateway also acts as a server, providing a cached superset of all the IOCs' values to public EPICS clients. 
In Figure~\ref{stats}, we see that the number of channels active in the current cache is 15692 and the number of channels that will be held in the cache for up to two more hours is 17854. 

The server is receiving requests and responding to them at the rate of 1343.56 channels per second. The software was built without the flags to monitor the number of file descriptors in use, so this explains "FDs: 0".

Plotted against time (Figure~\ref{graph}), we see that the number of alive and active channels is near constant and that there are small fluctuations in the event and post rates.
 
The dual network interface card (NIC) servers required a special EPICS configuration to avoid seeing ambiguous replies to their request for data (since they could access both the IOCs directly and the CA gateway for values). 

The BaBar private network is not routed, so IOCs are hidden from the
public internet address space.
 
IOCs use security access files to define access security groups and rules
to specify which users can carry out which actions. For example, the
access security group ``dchexpert'' may contain a list of users who are
allowed to turn on the high voltage for the drift chamber.
A security access file for the gateway was constructed from the sum of the individual
IOC access security files. The userid running the gateway processes had 
to be added to the individual IOC's security rules since requests to do secure
transactions at the IOC level were being made by the gateway process
and not the expert user, if initiated from the public network.

\section{Results}

The number of file descriptors decreased by 25\% and the CPU usage
decreased by 20--40\%, on average for the 17 IOCs. There were fewer
IOC hangs/disconnects which helped to improve BaBar's data-taking
efficiency. 

Internal IOC security is maintained by propagating the IOC security access
definitions to the gateway security definitions. External IOC security
is improved since IOCs are no longer publicly visible. 



\section{Appendix}
\begin{turnpage}
\begin{figure*}
\includegraphics[width=240mm]{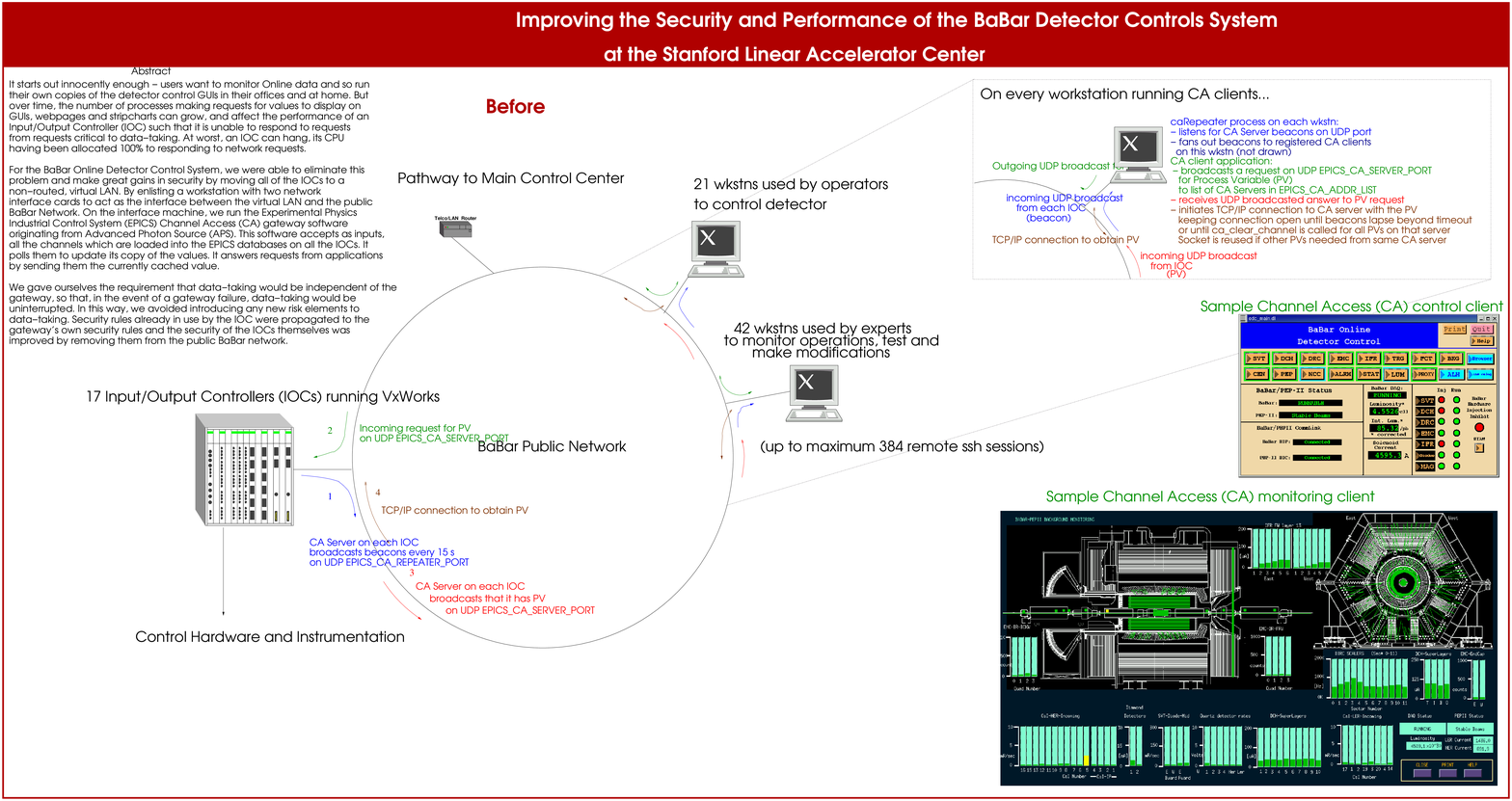}
\caption{CHEP03 poster.}
\label{top}
\end{figure*}
\end{turnpage}

\begin{turnpage}
\begin{figure*}
\includegraphics[width=240mm]{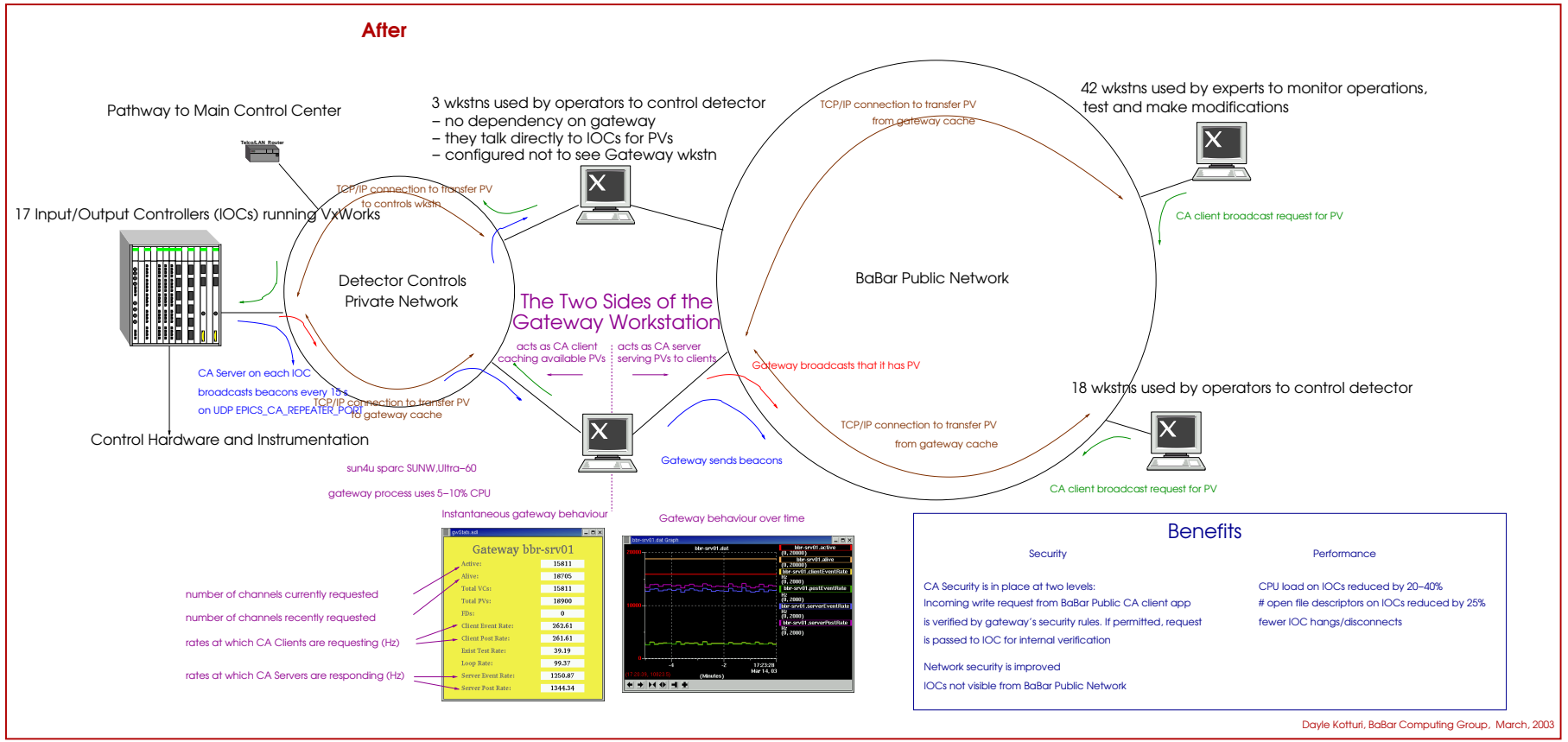}
\caption{CHEP03 poster contd.}
\label{bot}
\end{figure*}
\end{turnpage}

The poster displayed at CHEP 2003, in two halves (Figures~\ref{top} and
~\ref{bot}).

\end{document}